# Magnon-Phonon Coupling in Layered Antiferromagnet

Somsubhra Ghosh, Mainak Palit, Sujan Maity, and Subhadeep Datta [a]

*School of Physical Sciences, Indian Association for the Cultivation of Science,
2A & B Raja S. C. Mullick Road, Jadavpur, Kolkata - 700032, India*

[a] *Corresponding author: sspsdd@iacs.res.in*

Abstract. We present a fully analytical model of hybridization between magnon, and phonons observed experimentally in magneto-Raman scattering in van der Waals (vdW) antiferromagnets (AFM). Here, the representative material, FePS$_3$, has been shown to be a quasi-two-dimensional-Ising antiferromagnet, with additional features of spin-phonon coupling in the Raman spectra emerging below the Néel temperature ($T_N$) of approximately 120 K. Using magneto-Raman spectroscopy as an optical probe of magnetic structure, we show that one of these Raman-active modes in the magnetically ordered state is a magnon with a frequency of 3.7 THz (~ 122 cm$^{-1}$). In addition, one magnon band and three phonon bands are coupled *via* the magneto-elastic coupling evidenced by anti-crossing in the complete spectra. We consider a simple model involving only in-plane nearest neighbor exchange couplings (designed to give rise to a similar magnetic structure) and perpendicular anisotropy in presence of an out-of-plane magnetic field. Exact diagonalization of the Hamiltonian leads to energy bands which show that the interaction term gives rise to avoided crossings between the hybridized magnon and phonon branches. Realizing magnon-phonon coupling in two-dimensional (2D) AFMs is important for the verification of the theoretical predictions on exotic quantum transport phenomena like spin-caloritronics, topological magnonics, etc.

## INTRODUCTION

In magnetic insulators, transport of heat occurs due to interaction between phonons, impurities, and most importantly, collective excitation of spins, namely magnons. Moreover, contribution to thermal conductivity (K = $K_{magnon}$ + $K_{Phonon}$) of a magnetic system at low temperature due to magnons ($K_{magnon}$ ~ T$^2$) is higher than phonons ($K_{Phonon}$ ~ T$^3$). Consequently, the interaction between spin waves and lattice vibrations, magnon - phonon interaction, is important in order to understand relaxation processes occurring in magnetic crystals, particularly below magnetic transition temperature. For example, in body-center cubic (bcc) iron, for the equilibrium between the magnon and phonon, relaxation time increases as the system cools down from room temperature (~ 10$^{-9}$ s) to liquid-helium temperatures (~ 10$^{-6}$ s). Importantly, in applications like information processing, data transfer and logic circuits, distortion less and Joule-heat free magnon current through a magnetic semiconducting strip (e:g Y$_3$Fe$_5$O$_{12}$; YIG) in a field effect transistor (FET) geometry can be drastically modified depending on the ratio between magnetic and non-magnetic impurity-scattering potentials, in short, on magnon-lattice coupling. In this context, 2D vdW magnetic materials (CrI$_3$, Cr$_2$Ge$_2$Te$_6$ etc.) offer a solid-state platform for studying magnons in variety of dimensions in a single batch of sample; strictly 2D systems, quasi-2D systems or in bulk and exploring the limits of long-established Mermin-Wagner theorem with the effect of magnetic anisotropy. Further to the discovery of insulating FMs, 2D itinerant ferromagnetic order was recently explored in Fe$_3$GeTe$_2$, which has actuated experiments on electron transport in nanoelectronic devices for the manipulation of both, spin and charge, degrees of freedom in an atomically thin magnet.

The important role of collective excitation of spins, like magnons, in tunneling through thin magnetic insulators in spin-transistor device geometry has already been shown in vdW heterostructures. Interestingly, spin waves or magnons (frequency in GHz) in a conventional FM, under an external magnetic field, precess only counterclockwise. However, antiferromagnets possess two magnetic field induced split magnon modes (frequency in THz) with opposite polarization, otherwise degenerate. Currently, one of the major thrusts in magnon spintronics or magnonics is to search for the ultrafast antiferromagnetic magnons, preferably at room temperature, with long propagation length. Moreover, to unravel the technological aspect, it is imperative to investigate coupling between lattice dynamics (phonons) and spin waves, namely magnetoelastic waves, in antiferromagnets.

Here, we report theoretical analysis of magnetic field induced hybridization between magnon and phonons in 2D vdW AFM FePS$_3$ [1,2]. Advantages of choosing FePS$_3$ over thin film AFMs, Cr-based halides and MPS3 (where M = transition metals) are: (i) robustness of transition temperature ($T_N$ ~ 120 K) in the atomic layer limit [1], (ii) higher magnon transition temperature ($T_M$ ~ 60 K in 3 layers) in few layers withstanding quantum fluctuations in 2D [1], (iii) magnons with higher frequency (~ 3.7 THz) than conventional AFM insulators like YIG, Cr$_2$O$_3$, and Fe$_2$O$_3$, (iv) extremely air stable (over a period of 6 months) even a few layers. Coupling between one magnon band and nearby three phonon bands can be identified from the anti-crossing in the magneto-Raman spectra at 4K [2]. Adopting a simple model which involves in-plane nearest neighbor exchange couplings complying with the spin texture and perpendicular anisotropy in presence of an out-of-plane magnetic field. Exact diagonalization of the Hamiltonian leads to energy bands which show that the interaction term gives rise to avoided crossings between the hybridized magnon and phonon branches, as observed in the experiment. Our model can be applied equally to explain the experimental data of other FM/AFM materials with suitable exchange parameters considering their magnetic structure and thus provides a means to analyze cooperative excitations involving spin, lattice in magnetic insulators.

## EXPERIMENTAL DATA AND THEORETICAL MODEL

The low-temperature (4 K) magneto-Raman spectroscopy on targeted FePS$_3$ flake (lateral dimension 15 μm) were performed *via* backscattering with the laser excitation at 514 nm with on-sample power 400 μW [2]. Magnetic fields were then applied perpendicular (Faraday configuration)/parallel (Voigt configuration) to the plane of the substrate to record the evolution of Raman spectrum and varied slowly from B = -20 T to B = 30 T (see Figure 1), as explained in our previous report [2]. Experimental results concerning the hybrid magnon-polaron mode has been discussed in detail by Vaclavkova *et al.*, corroborated by Liu *et al.* [2-4]. The objective of this study is to create an analytical framework to understand the experimental result.

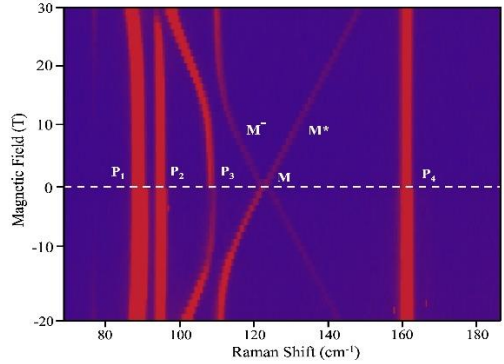

FIGURE 1. Low-temperature ($T$ = 4K) magneto-Raman scattering: Characteristic phonon modes are recorded *via* μ-Raman scattering measurement while magnetic field is swept from -20 T to +30 T. the phonon modes (P1, P2, P3 and P4) and magnon mode M at (122 cm$^{-1}$) can be identified from the color map of Raman spectral intensity. Avoided crossing around ± 10 T signifies magnon-phonon hybridization, as reported in [2].

Analytical treatment: Experimental evidence suggests that FePS$_3$ has a layered structure having ferromagnetic chains of Fe$^{2+}$ (S = 2) coupled antiferromagnetically to each other within a layer, with a very small interlayer coupling existing between them [4,5]. Such a magnetic structure on a honeycomb lattice is known to be stable for certain restricted values of the nearest neighbor, next nearest neighbor and third nearest neighbor couplings [6, 7]. Furthermore, it exhibits a Ising-like behavior with spins pointing out of the plane, owing to the presence of a strong z-axis anisotropy. We consider a simpler effective model here, involving only in-plane nearest neighbor exchange couplings (designed to give rise to a similar magnetic structure) and a z-axis anisotropy, as shown in Fig 2(a). In the presence of an out-of-plane magnetic field, $\vec{h} = h\hat{z}$, the magnetic Hamiltonian reads:

$$H_{mag} = -J\sum_{r\in A}\left(\vec{S_r}\cdot\vec{S_{r+\delta_1}} + \vec{S_r}\cdot\vec{S_{r+\delta_2}} - \vec{S_r}\cdot\vec{S_{r+\delta_3}}\right) - \Delta\sum_{r\in A,B}(S_r^z)^2 - H\sum_{r\in A,B}S_r^z \tag{1}$$

The exchange coupling, $J > 0$ is taken to be site-independent, but bond-dependent, such that two of the three nearest neighbors are ferromagnetically coupled, whereas the third couples antiferromagnetically. $\Delta > 0$ represents the anisotropy coefficient, while H = μ$_B$gh gives the Zeeman coupling to the external field, with μ$_B$ being the Bohr magneton and *g* the Landé g-factor. It can be shown that the classical ground state of this model Hamiltonian agrees with that depicted in Fig. 2(a) for H < JS and above a small threshold value of the anisotropy parameter, Δ. To study the low-lying magnon excitations ($\langle n_{mag}\rangle \ll S$), we consider the non-Bravais lattice to be composed of four interpenetrating magnetic sublattices as shown in Fig. 2(b) and apply the Holstein-Primakoff transformations:

Sublattice $a_1(b_1)$: $S_j^z = -S + b_{j,A(B)}^\dagger b_{j,A(B)}$, $S_j^+ = b_{j,A(B)}^\dagger \sqrt{2S - b_{j,A(B)}^\dagger b_{j,A(B)}}$, $S_j^- = \sqrt{2S - b_{j,A(B)}^\dagger b_{j,A(B)}}\, b_{j,A(B)}$

Sublattice $a_2(b_2)$: $S_j^z = S - a_{j,A(B)}^\dagger a_{j,A(B)}$, $S_j^- = a_{j,A(B)}^\dagger \sqrt{2S - a_{j,A(B)}^\dagger a_{j,A(B)}}$, $S_j^+ = \sqrt{2S - a_{j,A(B)}^\dagger a_{j,A(B)}}\, a_{j,A(B)}$

Retaining terms which are at most quadratic in the operators $a_{j,A(B)}$ and $b_{j,A(B)}$ (and their conjugates), the Hamiltonian $H_{mag}$ can be diagonalized using Bogoliubov transformation to give

$$H_{mag} = \sum_k \left(\epsilon_{k,m}^1 \alpha_{1,k}^\dagger \alpha_{1,k} + \epsilon_{k,m}^2 \alpha_{2,k}^\dagger \alpha_{2,k} + \epsilon_{k,m}^3 \beta_{1,-k}^\dagger \beta_{1,-k} + \epsilon_{k,m}^4 \beta_{2,-k}^\dagger \beta_{2,-k}\right) + \text{const} \tag{2}$$

where the sum over $k$ runs over the first magnetic Brillouin zone as shown in the inset of Fig. 2(b). There are four magnon branches with dispersion relations:

$$\epsilon_{k,m}^{1(2)} = H + \sqrt{A^2 + |B|^2 - |C|^2 \mp \sqrt{4A^2|B|^2 + (B^*C - BC^*)^2}}$$

$$\epsilon_{k,m}^{3(4)} = -H + \sqrt{A^2 + |B|^2 - |C|^2 \mp \sqrt{4A^2|B|^2 + (B^*C - BC^*)^2}} \tag{3}$$

where $A = 3JS + 2\Delta S$, $B = -2JS \exp\left(-\frac{ik_y a}{2}\right)\cos\left(\frac{\sqrt{3}}{2}k_x a\right)$ and $C = JS \exp(ik_y a)$.

To study the lattice dynamics, we consider a model in which the nearest neighbor $Fe^{2+}$ ions are elastically coupled to each other by an effective spring constant $K$, such that the elastic Hamiltonian reads:

$$H_{ph} = \frac{1}{2M}\sum_{r \in A,B}(p_r^z)^2 + \frac{1}{2}K\sum_{r \in A}\sum_\delta (u_r^z - u_{r+\delta}^z)^2 \tag{4}$$

where $M$ is the ion mass and $p_r^z = M\dot{u}_r^z$ is the canonically conjugate momentum. We have dealt with the out-of-plane phonon polarization only, as this is the one which, to leading order, couples to the magnons in the magnetoelastic coupling term treated later in this section. Considering two consecutive $Fe^{2+}$ ions in two crystallographically inequivalent sites $A$ and $B$ as forming a basis, we obtain an acoustic phonon branch and an optical phonon branch in the reduced Brillouin zone. The diagonalized elastic Hamiltonian has the form,

$$H_{ph} = \hbar \sum_k \left(\omega_{k,p}^- \eta_k^\dagger \eta_k + \omega_{k,p}^+ \zeta_k^\dagger \zeta_k\right) + \text{const} \tag{5}$$

Where $\omega_{k,p}^\pm = \sqrt{\frac{K}{M}(3 \pm |G(k)|)}$, with $G(k) = 2\exp\left(-i\frac{k_y a}{2}\right)\cos\left(\frac{\sqrt{3}}{2}k_x a\right) + \exp(ik_y a)$. The form of the magnetoelastic energy term is derived by following the procedure outlined in Ref [8-10]. A Taylor series expansion of the anisotropy energy in the elastic strain components and retention of leading order terms, consistent with the symmetry of the lattice and the magnetic ground state, gives for the magnetoelastic energy:

$$H_{coupled} = \kappa \sum_{r \in A,B}\sum_\delta S_r^z (\vec{S_r} \cdot \hat{\delta})(u_r^z - u_{r+\delta}^z) \tag{6}$$

where $\kappa$ is the magnetoelastic coupling constant and $\hat{\delta}$ are unit vectors connecting nearest neighbors, as shown in Fig. 2(a).

Comparison with experimental data: The total Hamiltonian of the coupled system has the form:

$$\mathcal{H} = H_{mag} + H_{ph} + H_{coupled} \tag{7}$$

This quadratic Hamiltonian can again be diagonalized using a Bogoliubov transformation to yield the energy levels of the system. Exact diagonalization of $\mathcal{H}$ for a given magnetic field leads to energy bands exhibiting avoided crossings between the magnon and phonon branches due to the presence of the magnon-phonon interaction.

Alternatively, this avoided crossing can be detected experimentally as a function of the applied field by focusing on one particular point in the Brillouin zone. Since magneto-Raman spectroscopy can access the $\Gamma$ point, we set $k_x a = k_y a = 0$ in our numerics. Fig. 2 (c) and 2(d) show the comparison between experimental data (solid dots) and our model predictions (continuous lines). The hybridization between the $M_-$ magnon branch and the $P1$ phonon branch is clearly visible. The $M_+$ branch does not show appreciable coupling to any of the phonon branches. Also, there are additional phonon branches (P2, P3) which do not emerge from this simplified model dealing only with the vibrations of $Fe^{2+}$ ions. The absence of additional magnon branches at higher energies in the experimental results might be explained in terms of selection rules, which falls outside the scope of our effective model. An estimate of the model parameters can be obtained by comparing the magnitude of this avoided crossing with the experimental data from Fig. 2. The best fit parameters of the model read: $J = 2.20$ meV, $\Delta = 2.50$ meV, $\hbar\sqrt{K/M} = 5.75$ meV and $\kappa = 115$ meV/nm. The interaction term being proportional to $S_z$ (to leading order), acts as an effective magnetic field applied in the $-z$ direction, which serves to lift the $Z2$ symmetry of the Hamiltonian $\mathcal{H}$ at zero field. This is manifested in the theoretical plot of figure Fig. 2(c) and 2(d) by the appearance of the splitting of magnon bands even at zero field. We would like to point out that such a splitting of the magnon branches at zero magnetic field has been observed experimentally in [2] with increased resolution. This can also be argued using the irreducible representation s (irreps) of the magnetic point group $2'/m$ to which $FePS_3$ belongs [5,6]. As shown in the next subsection, both the irreps of this group are non-degenerate. Hence any magnon mode within the crystal at zero field must be non-degenerate.

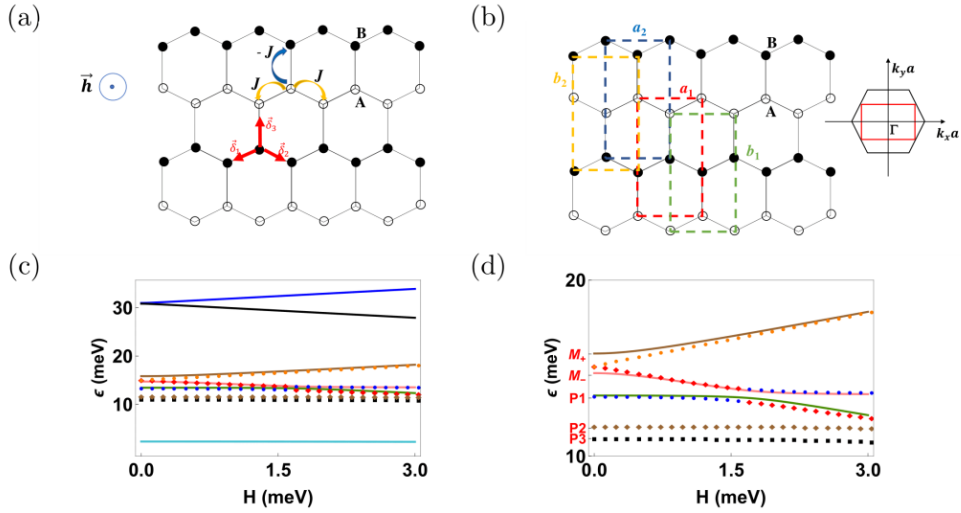

FIGURE 2. (a) Magnetic structure of a single layer of $Fe^{2+}$ ions in $FePS_3$. The black filled circles represent spins pointing out of the plane of the paper, while the white hollow circles represent spins pointing into the plane of the paper. Two of the nearest neighbors are ferromagnetically coupled, whereas the third is coupled antiferromagnetically. The two crystallographically inequivalent sites are labelled as A and B. For a A-type site, the nearest neighbor bonds are given by the vectors, $\vec{\delta_1} = a\left(-\frac{\sqrt{3}}{2}\hat{x} - \frac{1}{2}\hat{y}\right)$, $\vec{\delta_2} = a\left(\frac{\sqrt{3}}{2}\hat{x} - \frac{1}{2}\hat{y}\right)$, and $\vec{\delta_3} = a\hat{y}$, where $a$ is the lattice constant. The magnetic field $\vec{h}$ points out of the plane of the paper. (b) The structure can be interpreted to made up of 4 interpenetrating magnetic sublattices. Inset: The first magnetic Brillouin zone (in red) and the first crystallographic Brillouin zone (in black). (c) Comparison of theoretical results (continuous lines) and experimental results (dots) for the variation of the energy bands with the applied field. Fitted parameters are given in the text. (d) shows a zoomed-in portion of (c), with the different magnon and phonon branches marked.

The magneto-Raman spectra of $FePS_3$ studied in [2] showed that the $M_-$ magnon branch is co-polarized, whereas the $M_+$ branch is predominantly cross-polarized. To address this observation, we resort to the study of the magnetic point group of $FePS_3$. The character table of the concerned magnetic point group [11]:

| $2'/m$ | $E$ | $\sigma_h$ | $TC_2$ | $Ti$ |
|---|---|---|---|---|
| A' | 1 | 1 | $e^{i\theta}$ | $e^{i\theta}$ |
| A" | 1 | -1 | $e^{i\varphi}$ | $e^{-i\varphi}$ |

where $T$ denotes the time reversal operation. The Raman tensors corresponding to the irreducible representations of this group are given as [12]: $R_{A'} = \begin{pmatrix} A & B & 0 \\ D & B & 0 \\ 0 & 0 & I \end{pmatrix}$ and $R_{A''} = \begin{pmatrix} 0 & 0 & C \\ 0 & 0 & F \\ G & H & 0 \end{pmatrix}$

In the presence of a magnetic field along the $c$-axis of the compound, a linear correction may be expected, which can be ascertained from symmetry arguments viz.

$$R_{A'} = \begin{pmatrix} A & B & 0 \\ D & B & 0 \\ 0 & 0 & I \end{pmatrix} + B_z \begin{pmatrix} 0 & 0 & C \\ 0 & 0 & F \\ G & H & 0 \end{pmatrix} \quad (8)$$

Since $B_z$ flips sign under $\sigma_h$, but objects transforming under $A'$ doesn't, therefore, the tensor multiplying $B_z$ must be of $A''$ type. The experimental geometry is however of back-scattering type (incident and scattered light along z-axis). This implies that the electric fields in both the incident and the scattered light will have components in the $x - y$ plane. Therefore, we do not expect any contribution from the correction terms, and it is the $A'$ representation that we deal with. The electric field vector in the incident left-circularly polarized light is given by $\epsilon_i = (1, -i, 0)^T$. Since the scattered light from the $M_-$ branch is co-polarized, i.e., it is predominantly left-circularly polarized, therefore, the electric field present in the scattered light also has the form $\epsilon_s = (1, -i, 0)^T$. This implies that the entries of the Raman tensor $R_{A'}$ corresponding to the $M_-$ branch satisfy $A - iB + iD + E \approx 0$. Charting similar argument for the cross-polarized $M_+$ branch, it follows that for the $M_+$ branch, $A + iB + iD - E \approx 0$.

## CONCLUSION

In conclusion, we have proposed an effective model to explain the hybridization between the magnon and phonon branches as observed in the magneto-Raman spectra of the compound FePS$_3$ [2]. We have estimated our model parameters by comparing the magnitude of the avoided crossing obtained from our model with that obtained in experiments. In addition, by analyzing the polarization characteristics of the incident and scattered lights from the magnon branches, we have predicted relations between the entries of the Raman tensors of this compound.

## ACKNOWLEDGMENTS

Authors acknowledge the support from LNCMI, Grenoble for the magneto-Raman scattering. A part of the data presented in the Fig. 1 is published in the ref [2]. Special thanks to Dr. Marek Potemski and Dr. Clément Faugeras. S.D. would like to acknowledge Prof. K. Sengupta for fruitful discussion. S.D. also like to acknowledge DST-SERB Grant No. CRG/2021/004334. S.G. acknowledges CSIR for the fellowship. S.M. is grateful to DST-INSPIRE for his fellowship.